\documentclass[journal=jpccck,manuscript=letter,layout=singlecolumn]{achemso}

\usepackage[version=3]{mhchem} 
\usepackage[T1]{fontenc}       
\usepackage{siunitx} 

\author{Mit H. Naik}
\author{Indrajit Maity}
\author{Prabal K. Maiti}
\author{Manish Jain}
\email{mjain@iisc.ac.in}
\affiliation[Indian Institute of Science]
{Center for Condensed Matter Theory, Department of Physics, Indian Institute of Science, Bangalore 560012, India}

\title[]
{
Kolmogorov-Crespi Potential For Multilayer Transition Metal Dichalcogenides: 
Capturing Structural Transformations In Moir\'e Superlattices.
}


\begin{document}

%
%
%
%
%

\begin{abstract}
We develop parameters for the interlayer Kolmogorov-Crespi (KC) potential to study structural features of 
 four transition metal dichalcogenides (TMDs): MoS$_2$, WS$_2$, 
MoSe$_2$ and WSe$_2$.
We also propose a mixing rule to extend the parameters to their heterostructures.   
Moir\'e superlattices of twisted bilayer TMDs
have been recently shown to host shear solitons, topological point defects and 
ultraflatbands close to the valence band edge. Performing structural relaxations at the 
DFT level is a major bottleneck in the
study of these systems. We show that the parametrized KC potential can be used to obtain 
atomic relaxations in good agreement with DFT relaxations. Furthermore, the moir\'e superlattices
relaxed using DFT and the proposed forcefield yield very similar electronic band structures. 
\end{abstract}

\section{Introduction}
The growing family of layered materials 
and the 'Lego set' \cite{Persp.Geim} of possible heterostructures 
is an attractive field of research \cite{Science.Nov,NatPhoton.Sun}. Layered materials 
are composed of two-dimensional (2D) atomic layers weakly 
held together by van der Waals (vdW) forces. 
The widely used approach to theoretically simulate 2D materials 
and their heterostructures is through vdW 
corrected density functional theory \cite{PR.Kohn,2D.Qian,JPCM.Arkady,PRM.Naik} (DFT). 
This quantum mechanics based approach is accurate but computationally intractable for applications that require 
large scale simulations of 2D materials with more than 
10,000 atoms. One such application is the study of moir\'e patterns in 2D materials. 

On introducing a small-angle twist between the two layers of a bilayer system leads to the formation of 
a large scale moir\'e superlattice (MSL) \cite{2D.Oleg,twist.Naik,PRB.Kang}. 
MSLs have interesting properties at the electronic as well as the structural level. 
The moir\'e pattern is composed of various local high-symmetry stackings. 
\cite{twist.Naik,2D.Oleg,PRB.Kang}
Structural reconstructions lead to the formation of shear strain solitons at stacking 
boundaries and topological point defects in twisted bilayer graphene (tBLG) 
\cite{NMat.Matthew,NL.Jose,2D.Oleg,PNAS.Alden,2D.Wijk}.
Unconventional superconductivity was recently observed in tBLG at a 'magic' \cite{PNAS.MacD} twist angle
due to the formation of ultraflatbands close to the 
Fermi level \cite{Nature.Cao,Nature.Cao2, PRB.Lischner, PRL.Wu}.
Recent DFT calculations show
that small-angle twisted bilayer TMDs can also host ultraflatbands, 
shear solitons and topological point defects \cite{twist.Naik}.  
However, angles smaller than 3.5$^\circ$
could not be explored in this study \cite{twist.Naik} due to higher computational cost.
Obtaining the relaxed coordinates of the atoms in the MSL is a major bottle-neck in these
calculations.

Classical force-field based methods have been used to replace expensive DFT calculations 
to study the structural properties of 2D materials 
\cite{Nanosc.Zhu,arxiv.Ouyang,2D.Wijk,2D.Oleg,NL.Tosatti}. 
An important ingredient to classical force-field 
based methods is the proper modelling of the vdW interaction between layers. 
\cite{JCP.Oded,JCTC.Oded,PRL.Marom} 
The Lennard-Jones (LJ)
model has been widely used to account for the interlayer vdW interactions. 
Sliding one layer of a 2D material with respect to the other leads to 
different stacking configuration of the atoms, which have different binding energies. This stacking 
dependence of the binding energy is 
not captured by the LJ model. Tribology \cite{ACSNano.Feng,Carbon.Zhai} and the study of moir\'e patterns in 
2D materials are particularly sensitive to the stacking dependence 
\cite{JPCL.Oded,PRB.KC}.

To overcome this drawback, the Kolmogorov-Crespi\cite{PRB.KC,PRL.KC} (KC) potential has been developed for bilayer graphene. 
This potential includes an additional stacking dependent term coupled with the short-range interaction.
It has been successfully applied to graphene \cite{2D.Oleg,PRL.Fasolino, PRB.Choi}
and BN \cite{JPCC.Oded} to explore solitons in moir\'e patterns \cite{2D.Oleg,2D.Wijk}, 
dislocations, shear modes, self-retraction \cite{PRL.Zheng,PRB.Popov,PRL.Wang} and lubricity of graphene flakes \cite{PRB.Koren,PRB.Koren2,JPCM.Wijk,Carbon.Zhai}
and carbon nanotubes \cite{NL.Tosatti,RSCA.Xia,JPCC.Oz,PRB.KC,Carbon.Zhai}, 
etc. However, it
has been restricted to these materials and 
not been extended to the vast family of transition metal 
dichalcogenides, and their heterostructures \cite{CMS.Polcar,PRB.Levita}. 
We note that other methods like the continuum approach \cite{arxiv.Carr,arxiv.Massatt} and a 
Gaussian-LJ \cite{JAP.Jiang} method have been developed to treat shear in bilayer
 MoS$_2$.


In this work, we parametrize the KC potential for multilayers of 
four transition metal dichalcogenides: MoS$_2$, MoSe$_2$, WS$_2$ and WSe$_2$. 
The Stillinger-Weber (SW) \cite{PRB.SW,Chap.SW} force field is used to model the constituent single
layers and the KC potential determines the interlayer interactions.
The KC parametrization is performed to fit the binding energy of the various bilayer 
stackings computed using DFT.
We also propose a mixing rule for these parameters to simulate heterostructures using these 
materials.
The layer breathing mode and shear mode frequencies computed using these potentials are in 
good agreement with experiments. 
We use the SW and fitted KC parameters to perform the structural relaxation of the 3.5$^\circ$ and 
56.5$^\circ$ twisted bilayer MoS$_2$. 
We find that the atomic relaxations using these forcefield calculations are in good 
agreement with relaxations performed using DFT. Furthermore, the electronic band structure computed using the 
forcefield relaxed structure is in good agreement with that computed using the DFT relaxed structure. 
Indicating that this method \emph{can replace} the computationally 
expensive DFT relaxation steps while studying the electronic 
properties of moir\'e superlattices.

\begin{figure}
  \includegraphics[scale=1.8]{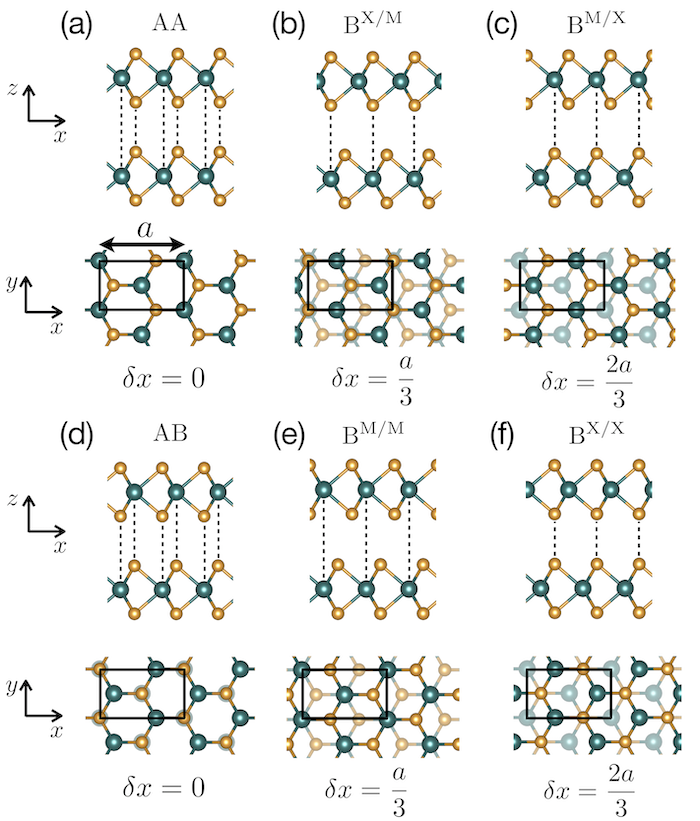}
  
 \caption{\label{fig1} 
  High-symmetry stackings of a bilayer TMD obtained by sliding the top layer 
  with repect to the bottom layer. The orthogonal unit cell is marked. 
  Starting with the AA stacking ($\delta x = \delta y = 0$) (a), sliding by $\delta x = a/3$ leads to the 
  B$^\mathrm{X/M}$ (b) stacking and sliding by $\delta x = 2a/3$ yields the B$^\mathrm{M/X}$ (c) stacking. 
  Similarly, starting with the AB (d) stacking yields B$^\mathrm{M/M}$ (e) and B$^\mathrm{X/X}$ (f) stackings.
}
\end{figure}

\begin{figure}
  \includegraphics[scale=2.32]{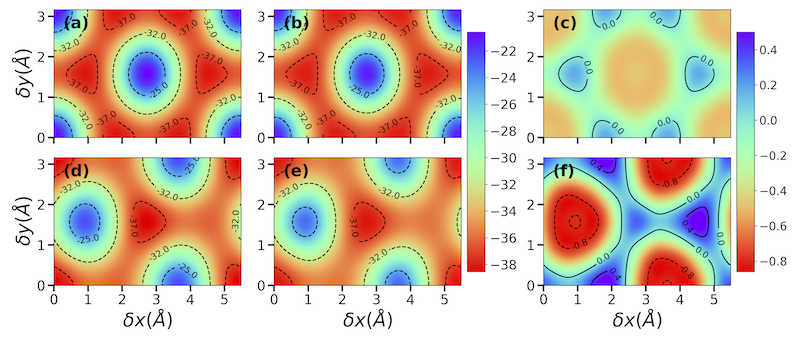}

 \caption{\label{fig2}
 Binding energy (BE) (in meV/atom) as a function of sliding the top layer of bilayer MoS$_2$ with respect to 
 the bottom layer along the $x$ and $y$ directions of the orthogonal unit cell. 
 The interlayer spacing is fixed at 6.35 \AA. The BE is computed as 
 $(\mathrm{E_{BL}-2E_{SL}})/\mathrm{N}$, where $\mathrm{E_{BL}}$ is
 the total energy of MoS$_2$ bilayer,
 $\mathrm{E_{SL}}$ is the energy of a monolayer and N is the number of atoms in the 
 bilayer system.
 (a) and (d) BE landscape obtained using van der Waals corrected DFT. 
 (b) and (e) BE landscape using Stillinger-Weber (SW) forcefield for intralayer interactions 
  and KC for interlayer interactions. 
 (c) and (f) Difference in the BE landscape computed using SW+KC and DFT.
  $(\delta x,\delta y) = (0,0)$ is the AA (AB) stacking for 
  (a), (b) and (c) ((d), (e) and (f)).
  }
\end{figure}

\begin{figure}
  \includegraphics[scale=1.08]{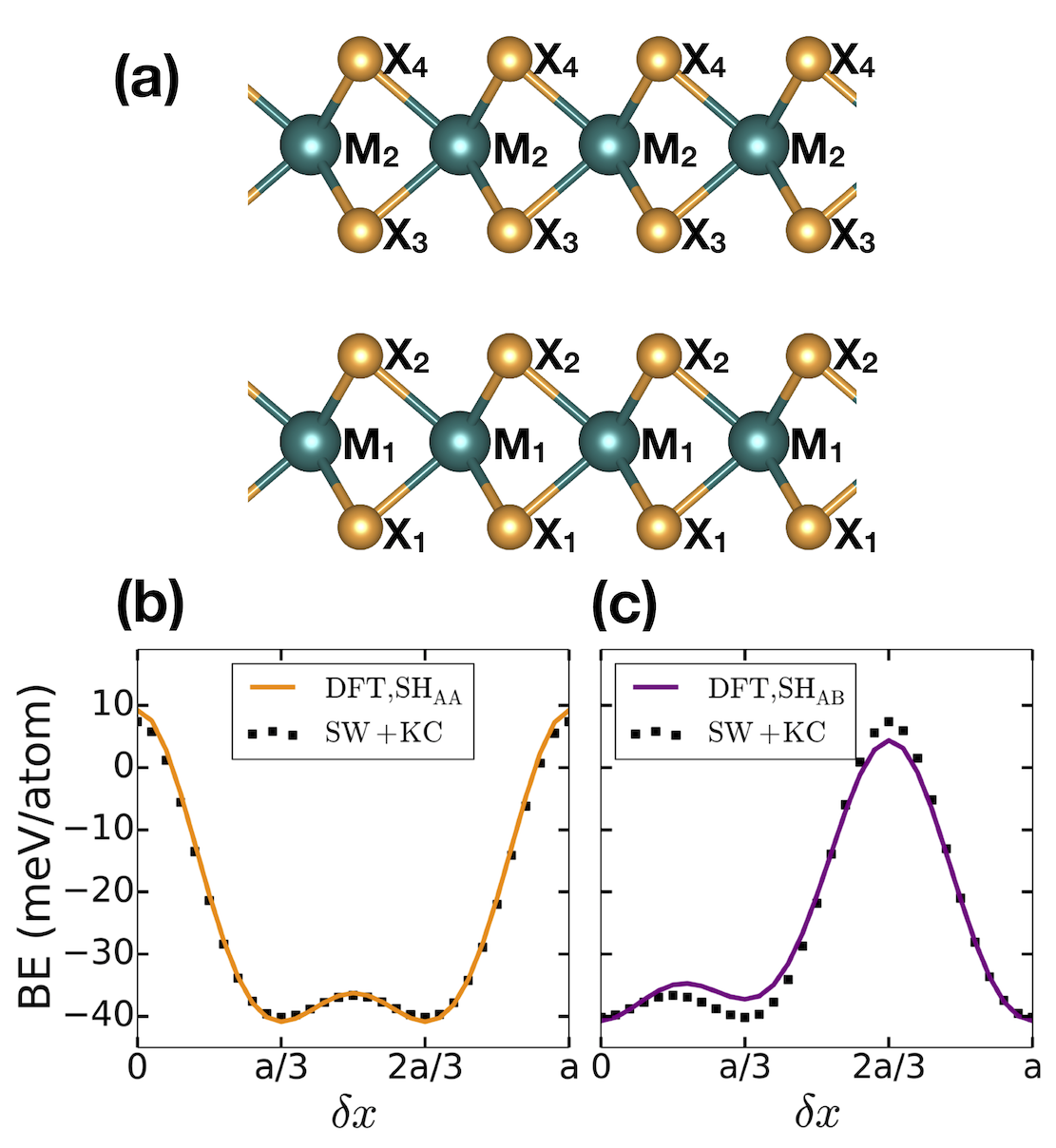}

 \caption{\label{fig3}
 (a) Bilayer of a TMD with labels for each layer and type of atom.
 (b) BE computed using DFT and SW+KC$^\mathrm{S-S}$ of bilayer MoS$_2$ as a
     function of shear starting with the AA stacking, $\mathrm{SH}_\mathrm{AA}$.
 (c) BE computed using DFT and SW+KC$^\mathrm{S-S}$ of bilayer MoS$_2$ as a
     function of shear starting with the AB stacking, $\mathrm{SH}_\mathrm{AB}$.
 In (b) and (c) the KC interaction is considered only between the S$_2$ and S$_3$ atoms of bilayer MoS$_2$.
 The interlayer spacing is fixed at 6 \AA. 
}
\end{figure}

\section{Computational Details}
The vdW corrected DFT calculations are carried out using the plane-wave pseudopotential 
package Quantum Espresso \cite{QE.Giannozi}.  
We use ultrasoft \cite{PRB.Vanderbilt} pseudopotentials and the 
local density approximation \cite{PRB.Perdew} (LDA) for the exchange-correlation 
functional. The van der Waals interactions are computed using the van der Waals 
density functional \cite{PRL.Thonhauser,PRB.Thonhauser} 
along with Cooper \cite{PRB.Cooper} exchange (vdW-DF-C09) \cite{RPP.Cooper}.  
We use a plane-wave energy cut-off of 50 Ry for the wavefunctions and a cut-off of 500 Ry 
for the charge density. The Brillouin zone is sampled with a 12$\times$12$\times$1 k-point mesh. 
The DFT calculations are performed with the optimized geometry of the cell for each TMD. 
The forcefield calculations are performed using the LAMMPS \cite{JCP.Plimpton,LAMMPS} package. 
The SW \cite{Chap.SW} potential 
is used for intralayer interactions and KC potential for interlayer interactions. The 
minimizations are performed using conjugate gradient method (force tolerance of 10$^{-6}$ 
eV/$\mathrm{\AA}$) as implemented in LAMMPS with the 
DFT lattice constant. Parametrization of the KC potential to fit the DFT binding energy
is performed with help of the Dakota \cite{UM.Dakota,Dakota} package. 
Commensurate moir\'e superlattices of twisted
bilayer MoS$_2$ are constructed using the Twister code \cite{twister}.
KC potential file and LAMMPS input files for the bilayer TMDs are available with the 
Twister package \cite{twister}.

\section{Kolmogorov-Crespi potential}
Fig. \ref{fig1} shows the various possible high-symmetry stackings of a bilayer TMD system obtained 
by sliding one layer with respect to the other \cite{twist.Naik}.  
The AA and AB stackings are obtained when atoms in 
the top layer are exactly above atoms in the bottom layer. 
In the case of AA stacking (Fig. \ref{fig1} (a)), similar atoms 
are on top of one another (M on M and X on X), while in AB stacking (Fig. \ref{fig1} (d)) 
dissimilar atoms are on top of one 
another (M on X and X on M). 
The other stackings are Bernal type (Fig. \ref{fig1} (b), (c), (e) and (f)), 
ie. one atom in the top layer is 
in the hexagonal cavity of the bottom layer. 
In our notation, B$^\mathrm{M/X}$ implies a Bernal (B)
stacking and that atom M in the top layer is directly above atom X in the bottom layer (Fig. \ref{fig1} (c)). 
On starting with the AA stacking, we obtain B$^\mathrm{X/M}$ and B$^\mathrm{M/X}$ on sliding 
one layer with respect to the other in the $x$ direction as shown in Fig. \ref{fig1}. We 
denote this shear by $\mathrm{SH}_\mathrm{AA}$.
Note that B$^\mathrm{X/M}$ 
and B$^\mathrm{M/X}$ are equivalent stackings. 
On sliding, starting with AB stacked layers, two non-equivalent high-symmetry
stackings are obtained: B$^\mathrm{M/M}$ and B$^\mathrm{X/X}$ (Fig. \ref{fig1}). 
We denote this shear by $\mathrm{SH}_\mathrm{AB}$.
There are thus five unique high-symmetry stackings 
possible with TMDs in the 2H-phase. 

We aim to model the binding energy (BE) landscape of the bilayer system as a function of shear between the two 
layers using classical forcefields. 
The BE landscape computed using vdW corrected DFT 
for bilayer MoS$_2$ with fixed interlayer spacing of 6.35 $\mathrm{\AA}$
 is shown in Fig. \ref{fig2} (a) and (d).
The Kolmogorov-Crespi is a reliable interlayer potential since it contains an
explicit registry dependent term. The form of the potential is given by:

\begin{equation}
  \label{eqn1}
    V_{ij} = e^{-\lambda (r_{ij} - z_{0})}V_{\rho} - A \left(\frac{r_{ij}}{z_{0}}\right)^{-6}
  \end{equation}

The potential, $V_{ij}$, is defined between atom $i$ in one layer and atom $j$ in the adjacent layer. 
The potential is set to zero after a cut-off radius, $r_{cut}$.
The KC potential includes a stacking dependent term, $V_{\rho}$, multiplying the 
short-range repulsive interaction.

\begin{equation}
V_{\rho} = [C + f(\rho_{ij}) + f(\rho_{ji})],
\end{equation}

where, $\rho_{ij}^2 = r_{ij}^2 - (\mathbf{n_i} \cdot \mathbf{r_{ij}})^2 $,
 $\rho_{ji}^2 = r_{ij}^2 - (\mathbf{n_j} \cdot \mathbf{r_{ij}})^2 $ and

\begin{equation} 
f(\rho) = e^{(-\rho/\delta)^2} \sum_{n=0}^{2} C_{2n} (\rho/\delta)^{2n}
\end{equation}

\begin{figure}
  \includegraphics[scale=0.88]{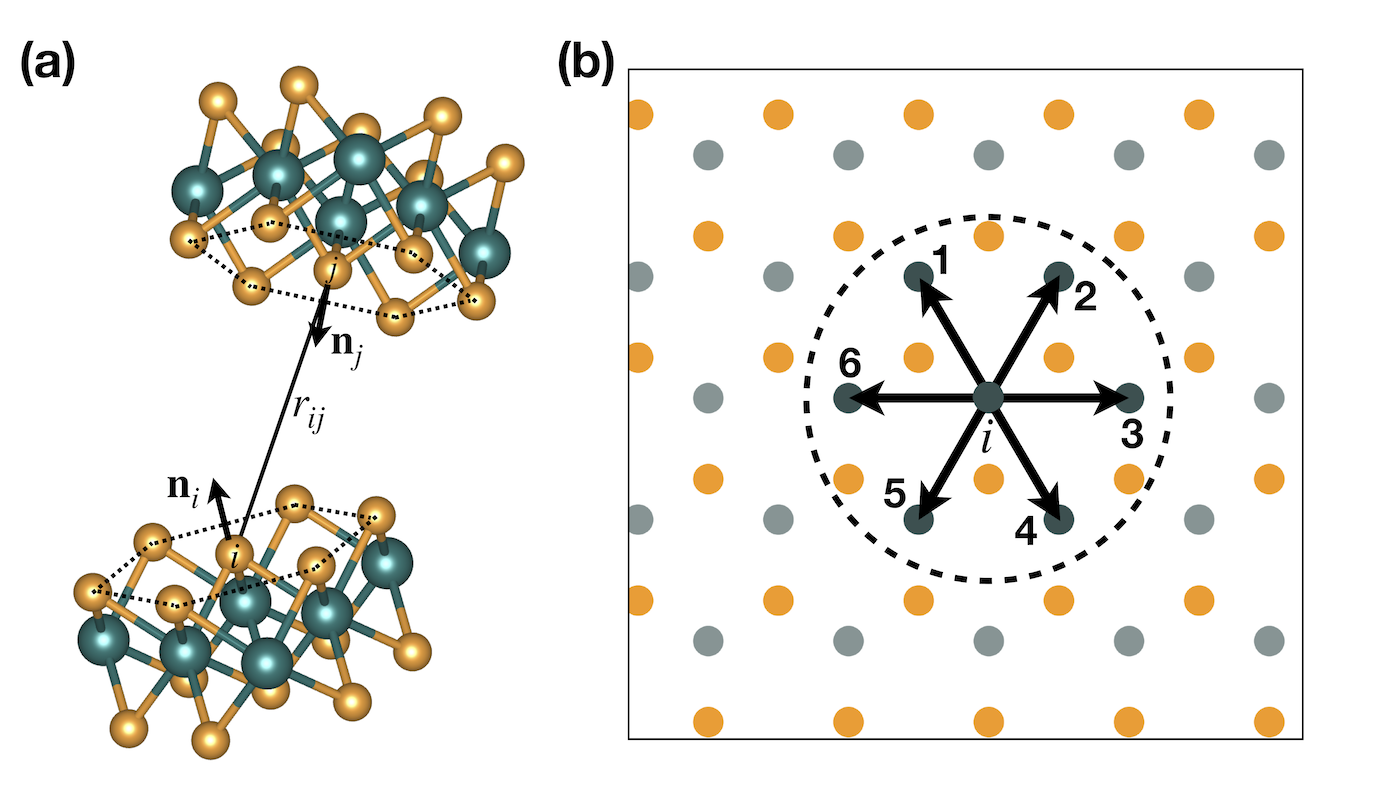}

 \caption{\label{figN}
 (a) Visual representation of the terms in the KC potential. The six nearest neighbours of the same type 
     as $i$ or $j$ used to construct the normal are connected using dotted lines.
 (b) Computation of local normal for each atom, $i$, depending on it's nearest-neighbor atoms of same type. For example, Mo atoms in $\mathrm{MoS_{2}}$ forms a triangular lattice with six nearest-neighbour Mo atoms, chosen using a predefined cut-off radius $r_{n}$. Six normals are computed using consecutive pairs of the 
 neighbour atoms: $\frac{\mathbf{r_{1i}}\times \mathbf{r_{2i}}}{|\mathbf{r_{1i}}\times \mathbf{r_{2i}}|}$, 
 $\frac{\mathbf{r_{2i}}\times \mathbf{r_{3i}}}{|\mathbf{r_{2i}}\times \mathbf{r_{3i}}|}$, 
 $\frac{\mathbf{r_{3i}}\times \mathbf{r_{4i}}}{|\mathbf{r_{3i}}\times \mathbf{r_{4i}}|}$, 
 $\frac{\mathbf{r_{4i}}\times \mathbf{r_{5i}}}{|\mathbf{r_{4i}}\times \mathbf{r_{5i}}|}$, 
 $\frac{\mathbf{r_{5i}}\times \mathbf{r_{6i}}}{|\mathbf{r_{5i}}\times \mathbf{r_{6i}}|}$ and 
 $\frac{\mathbf{r_{6i}}\times \mathbf{r_{1i}}}{|\mathbf{r_{6i}}\times \mathbf{r_{1i}}|}$. The local normal is finally computed after averaging over the six normals for $i$-th atom. 
}
\end{figure}

$V_{\rho}$ determines the energy barrier to shear one 
layer with respect to the other in a bilayer system. 
The KC potential thus consists of 8 parameters for each type of interaction. 
A taper function \cite{JCTC.Oded,JCP.Oded,arxiv.Wen,arxiv.Ouyang}
 is often used to ensure the interlayer potential goes to zero smoothly. Using a large
$r_{cut}$ = 14 \si{\angstrom} does not affect the results significantly in the absence of the taper function.
See Supplementary Information (SI) for more details.
$\mathbf{n_i}$ and $\mathbf{n_j}$ are surface normals at site $i$ and $j$, respectively. To compute the surface 
normal at the position of an atom $i$, we find the neighbours of the same type in a radius $r_n$ around
atom $i$. This is illustrated in Fig. \ref{figN}. $r_n$ is chosen to accommodate the six nearest neighbours of the 
same type. Six normals are then constructed from consecutive pairs of neighbouring atoms and averaged to obtain 
$\mathbf{n_i}$. See SI for more details. We refer to interactions 
that take into account the normals as KC-n.
A simplification to the present 
form of the potential can be introduced by setting $\mathbf{n_i} = \mathbf{n_j} = \hat{z}$. This 
approximation only works for multilayers whose bending leads to normals that do not deviate 
significantly from $\hat{z}$. For the case of twisted bilayers, we find this to be a good approximation 
since the normals are close to $\hat{z}$. We will refer to this approximation as KC-z.
This is discussed further in subsequent sections. 



\begin{figure}
  \includegraphics[scale=0.5]{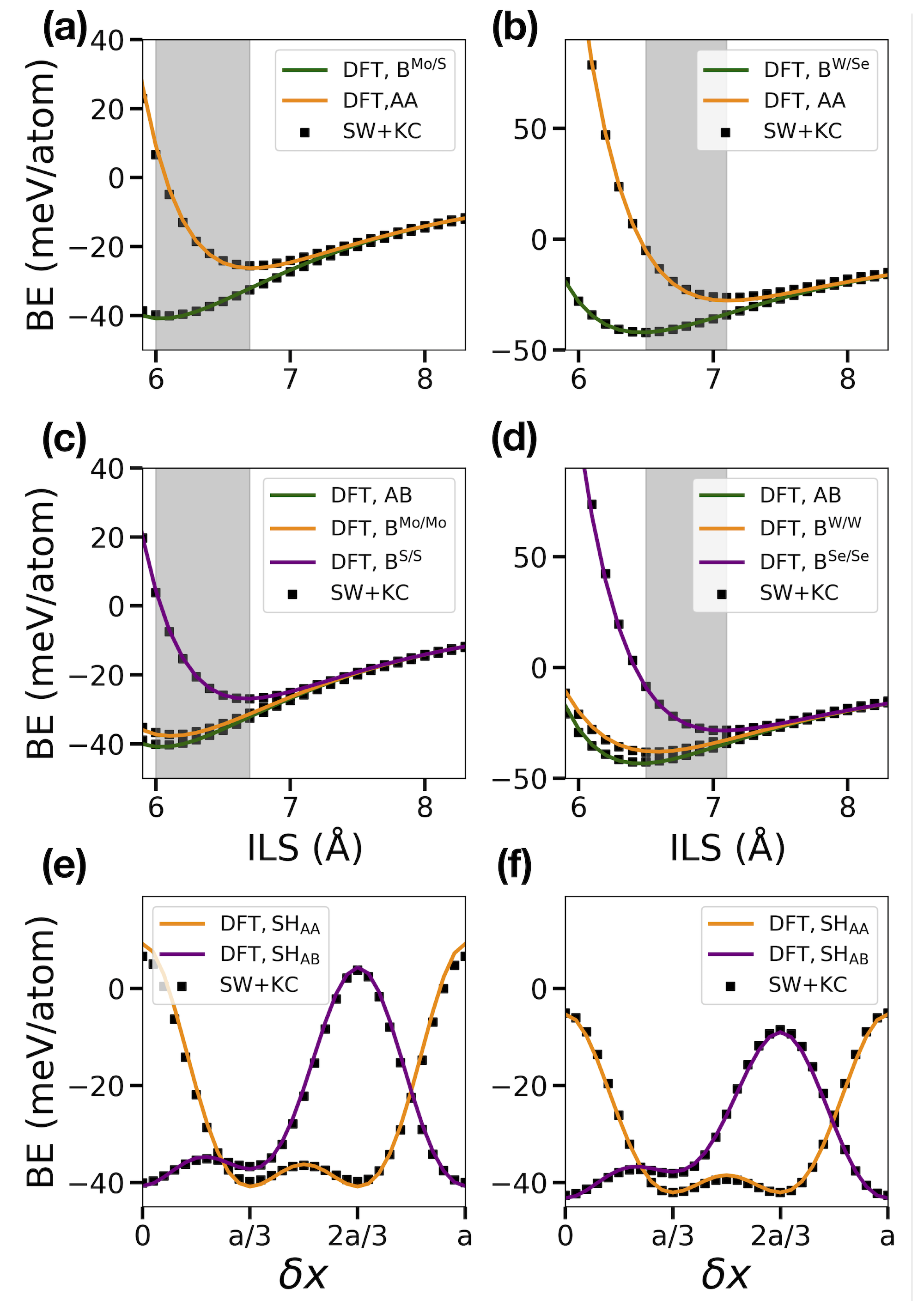}

 \caption{\label{fig5} 
 (a) and (c) ((b) and (d)) BE computed within DFT and using SW+KC 
 as a function of 
 interlayer spacing for the high-symmetry stackings in bilayer MoS$_2$ (WSe$_2$). 
 The shaded region marks the range of interlayer spacings in the bilayer system.
 (e) ((f)) BE computed within DFT and using SW+KC 
 as a function of
 shear, $\mathrm{SH}_\mathrm{AA}$ and $\mathrm{SH}_\mathrm{AB}$, in bilayer MoS$_2$ (WSe$_2$).  
 The ILS in (e) and (f) is fixed to 6.0 and 6.5 \AA.
}
\end{figure}

\begin{figure}
  \includegraphics[scale=0.5]{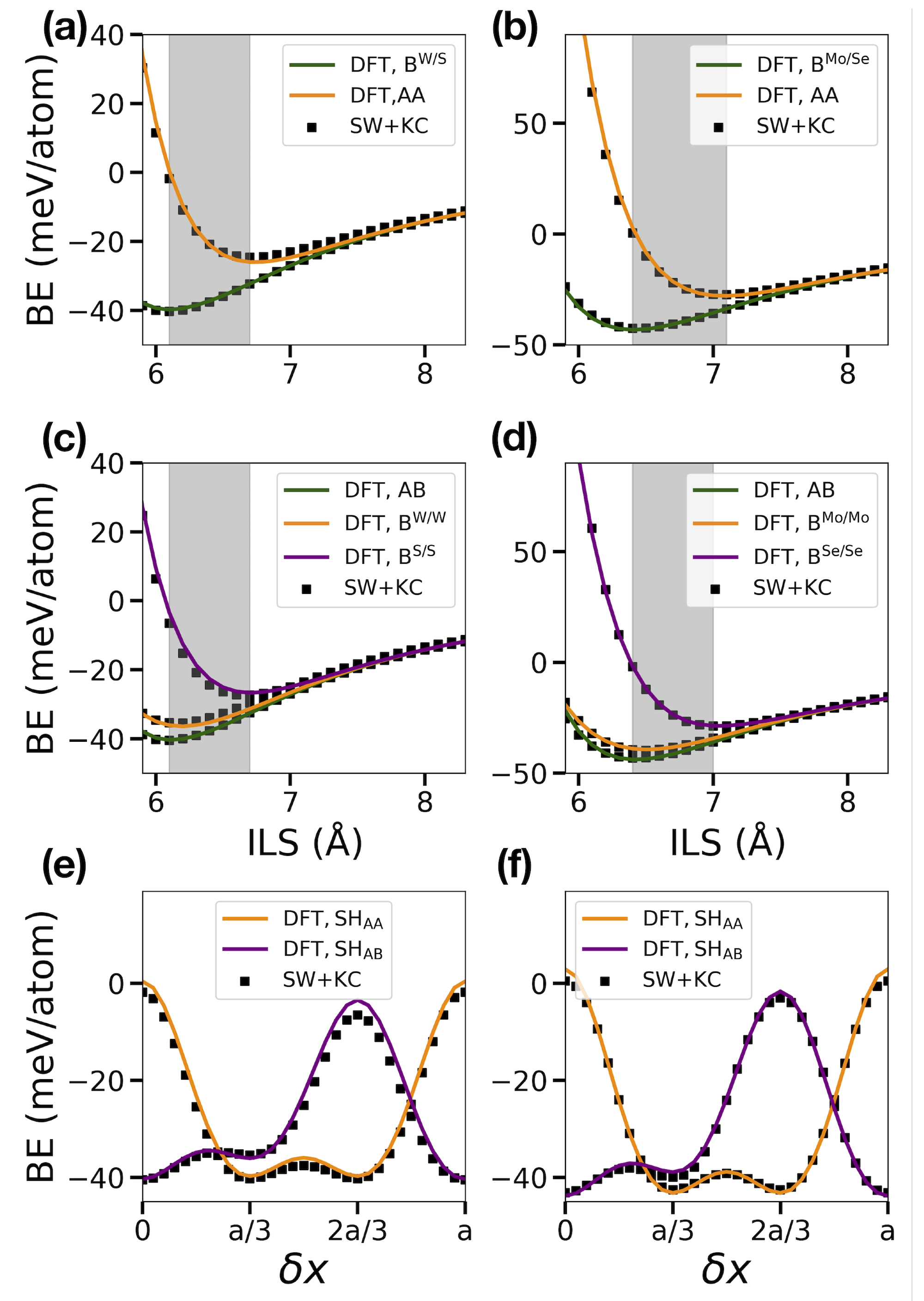}

 \caption{\label{fig6}
 (a) and (c) ((b) and (d)) BE computed within DFT and using SW+KC
 as a function of
 interlayer spacing for the high-symmetry stackings in bilayer WS$_2$ (MoSe$_2$).
 The shaded region marks the range of interlayer spacings in the bilayer system.
 (e) ((f)) BE computed within DFT and using SW+KC  
 as a function of
 shear, $\mathrm{SH}_\mathrm{AA}$ and $\mathrm{SH}_\mathrm{AB}$, in bilayer WS$_2$ (MoSe$_2$).
 The ILS in (e) and (f) is fixed to 6.0 and 6.4 \AA.
}
\end{figure}

Steric effects govern the relative energy of the stackings.
The stackings with chalcogen atoms in the top layer directly above chalcogen atoms in the 
bottom layer, ie. AA and B$^\mathrm{X/X}$, are unfavourable and highest in energy \cite{twist.Naik,PRB.Naik}. 
The other three stackings are relatively lower in energy. 
It would thus seem reasonable to consider the KC interactions between the X$_2$ and X$_3$  (Fig. \ref{fig3} (a)) 
atoms alone. On fitting the KC parameters of this interaction, KC$^\mathrm{X-X}$, to DFT BE 
leads to a good representation of 
shear starting with the AA stacking ($\mathrm{SH}_\mathrm{AA}$), as shown in Fig. \ref{fig3} (b).
But fails to completely represent the BE of shear starting with the AB stacking (Fig. \ref{fig3} (c)).
In particular, the 
difference in BE between AB and B$^\mathrm{M/M}$ stackings is not captured.
This is expected, because in the presence of only X$_2$-X$_3$ interactions, 
AB stacking is indistinguishable from the B$^\mathrm{M/M}$ stacking.  

To differentiate the BE of the AB and B$^\mathrm{M/M}$ stackings in $\mathrm{SH}_\mathrm{AB}$ we further 
introduce the M$_1$-X$_3$, M$_2$-X$_2$ and M$_1$-M$_2$ interlayer interactions 
(see Fig. \ref{fig3} (a)). We introduce these additional interactions and refit all the KC parameters. 
In the supplementary information we propose two other approaches to obtain the BE landscape 
with reasonable accuracy. The first includes only X$_2$-X$_3$ and M$_1$-M$_2$ interactions, 
KC$^\mathrm{X-X}_\mathrm{M-M}$, and the second approach includes only 
X$_2$-X$_3$, M$_1$-X$_3$ and M$_2$-X$_2$ interactions, KC$^\mathrm{X-X}_\mathrm{M-X}$. 
We have also fit KC$^\mathrm{X-X}_\mathrm{M-M}$ parameters including 
the taper function discussed above.
KC$^\mathrm{X-X}_\mathrm{M-M}$ and KC$^\mathrm{X-X}_\mathrm{M-X}$ parameters are provided and 
further discussed in the supplementary information. For the rest of the article, we will 
show the performance of the parametrized KC potential when
all interlayer interactions, X$_2$-X$_3$, M$_1$-X$_3$, M$_2$-X$_2$ and M$_1$-M$_2$, are 
included. These parameters are provided in Table 1. The BE landscape for MoS$_2$ 
with SW+KC is shown in Fig. \ref{fig2} (b) and (e). Fig. \ref{fig2} (c) and (f) 
show the deviation of the landscape computed using SW+KC
from DFT. 

\begin{table*}
\centering
\begin{tabular}{c@{\hskip 0.07in}c@{\hskip 0.07in}c@{\hskip 0.07in}c@{\hskip 0.07in}c@{\hskip 0.07in}c@{\hskip 0.07in}c@{\hskip 0.07in}c@{\hskip 0.07in}c}
\hline
\hline
  & \multicolumn{1}{c@{\hskip 0.07in}}{$z_0$}
& \multicolumn{1}{c@{\hskip 0.07in}}{C$_0$}
& \multicolumn{1}{c@{\hskip 0.07in}}{C$_2$}
& \multicolumn{1}{c@{\hskip 0.07in}}{C$_4$}
& \multicolumn{1}{c@{\hskip 0.07in}}{C}
& \multicolumn{1}{c@{\hskip 0.07in}}{$\delta$}
&  \multicolumn{1}{c@{\hskip 0.07in}}{$\lambda$} & \multicolumn{1}{c@{\hskip 0.07in}}{A}
\\
& (\AA) & (meV) & (meV) & (meV) & (meV) & (\AA) & (\AA$^{-1}$) & (meV)
         \\
\hline
S$_2$-S$_3$ &    3.857 & 7.074 & 2.624 & 0.024 & 24.859 & 0.982 & 2.762 & 55.637 \\
Se$_2$-Se$_3$  & 3.992 & 7.886 & 6.355 & -0.011 & 26.762 & 0.935 & 2.753 & 60.829 \\
Mo$_1$-S$_3$, Mo$_2$-S$_2$ & 4.049 & -2.718 & -2.719 & -0.325 & 16.459 & 0.616 & 2.564 & 14.239 \\
Mo$_1$-Se$_3$, Mo$_2$-Se$_2$ & 5.080 & -1.120 & -0.761 & -0.117 & 4.624 & 0.777 & 1.996 & 6.209 \\
W$_1$-S$_3$, W$_2$-S$_2$ &  4.331 & -1.344 & -1.370 & -0.162 & 8.372 & 0.587 & 2.082 & 7.078 \\
W$_1$-Se$_3$, W$_2$-Se$_2$ & 5.079 & -1.109 & -0.762 & -0.120 & 4.627 & 0.783 & 1.999 & 5.112  \\
Mo$_1$-Mo$_2$ &  9.236 & 1.471 & 1.126 & 0.131 & 1.928 & 1.183 & 0.924 & 3.439  \\
W$_1$-W$_2$ &   10.545 & 0.254 & 0.139 & -0.028 & 0.506 & 1.010 & 0.944 & 1.356 \\
\hline
\hline
\end{tabular}
\caption{ Parameters for interlayer Kolmogorov-Crespi interactions.
}
\end{table*}

\begin{table}
        \centering
        \begin{tabular}{*{3}{c}}
        \hline
        \hline
    \multicolumn{1}{c}{} & \multicolumn{2}{c}{LBM (cm$^{-1}$)} 
    \\
     &  \multicolumn{1}{c}{SW+KC} & \multicolumn{1}{c}{Expt.}
    \\
   \hline
   BLMoS$_2$  &  38.6  & 40 \cite{NL.Zhao}, 41.6 \cite{Chen_2015}   \\
    BLWS$_2$  &  30.5 & 33.8 \cite{Chen_2015}   \\
    BLMoSe$_2$   &  31.0 & 34.3 \cite{Chen_2015}, 29 \cite{cvd_tmd}  \\
      BLWSe$_2$ &  27.2  &  29.1 \cite{Chen_2015}, 27 \cite{cvd_tmd}  \\ 
   \hline
     \multicolumn{1}{c}{} & \multicolumn{2}{c}{SM (cm$^{-1}$)} \\

     &  \multicolumn{1}{c}{SW+KC$^\mathrm{X-X}_\mathrm{M-M}$} & \multicolumn{1}{c}{Expt.} \\ 
        \hline
   BLMoS$_2$    & 20.9 &  22 \cite{NL.Zhao}, 24.2 \cite{Chen_2015}  \\
    BLWS$_2$    & 18.0 & 19.6 \cite{Chen_2015}  \\
    BLMoSe$_2$  & 18.3 & 21 \cite{Chen_2015}, 18 \cite{cvd_tmd}  \\
      BLWSe$_2$ & 15.7 & 17.7 \cite{Chen_2015}, 17 \cite{cvd_tmd} \\
        \hline
        \hline

    \end{tabular}
     \caption{Comparison of the shear (SM) and layer breathing modes (LBM) of bilayer TMDs
        computed using SW+KC with experimental measurements.}
\end{table}

The KC potential parameters are obtained by fitting 
 the BE as a function of interlayer 
spacing (out-of-plane separation between M atoms of top and bottom layer) 
for the high-symmetry stackings and as a function of 
shear between the two layers, $\mathrm{SH}_\mathrm{AA}$ and $\mathrm{SH}_\mathrm{AB}$. 
The fitting to the shear is performed at the average 
interlayer spacing of the high-symmetry stackings. 
Fitting shear at the average interlayer 
spacing is sufficient to reproduce the shear at the minimum and maximum interlayer spacings as well.
Fig. \ref{fig5} (a), (c) and (e) compare the BE computed using SW+KC 
with the corresponding DFT
BE for bilayer MoS$_2$.  
Fig. \ref{fig5} (b), (d) and (f) compares SW+KC BE with DFT for bilayer WSe$_2$.
Fig. \ref{fig6} similarly shows the performance of the KC parameters for bilayer 
WS$_2$ and MoSe$_2$.
Furthermore, the shear and layer breathing modes \cite{Maity_PRB_2018} of the bilayer 
TMDs computed using SW+KC are in good agreement with 
experimental measurements (Table 2). 
See supplementary information for the performance of the other set of parameters: 
KC$^\mathrm{X-X}_\mathrm{M-M}$ and KC$^\mathrm{X-X}_\mathrm{M-X}$.

\section{Heterostructures: mixing rule}

Mechanical exfoliation \cite{NS.Wang} and chemical vapour deposition \cite{NL.Tongay,NL.Neaton,NL.Gong} 
techniques can be 
used to construct vertical stacks \cite{NRM.Liu} of different TMDs. 
These heterostructures 
generally form a Type II heterojunction \cite{NS.Wang}, which makes them suitable for applications in
nano- and opto-electronics\cite{NRM.Liu}. The KC parameters developed above for the bilayer TMDs can be 
extended to heterostructures by the use of a simple mixing rule. 
The cross-interaction parameters for S-Se are obtained by taking the
arithmetic mean (AM) of S-S and Se-Se parameters. Similarly the Mo-W interaction are computed 
as an AM of Mo-Mo and W-W parameters. 
We tried other mixing rules like the geometric mixing rule and 
find that the AM works best to reproduce the DFT BE. 

\begin{figure}
 \includegraphics[scale=0.92]{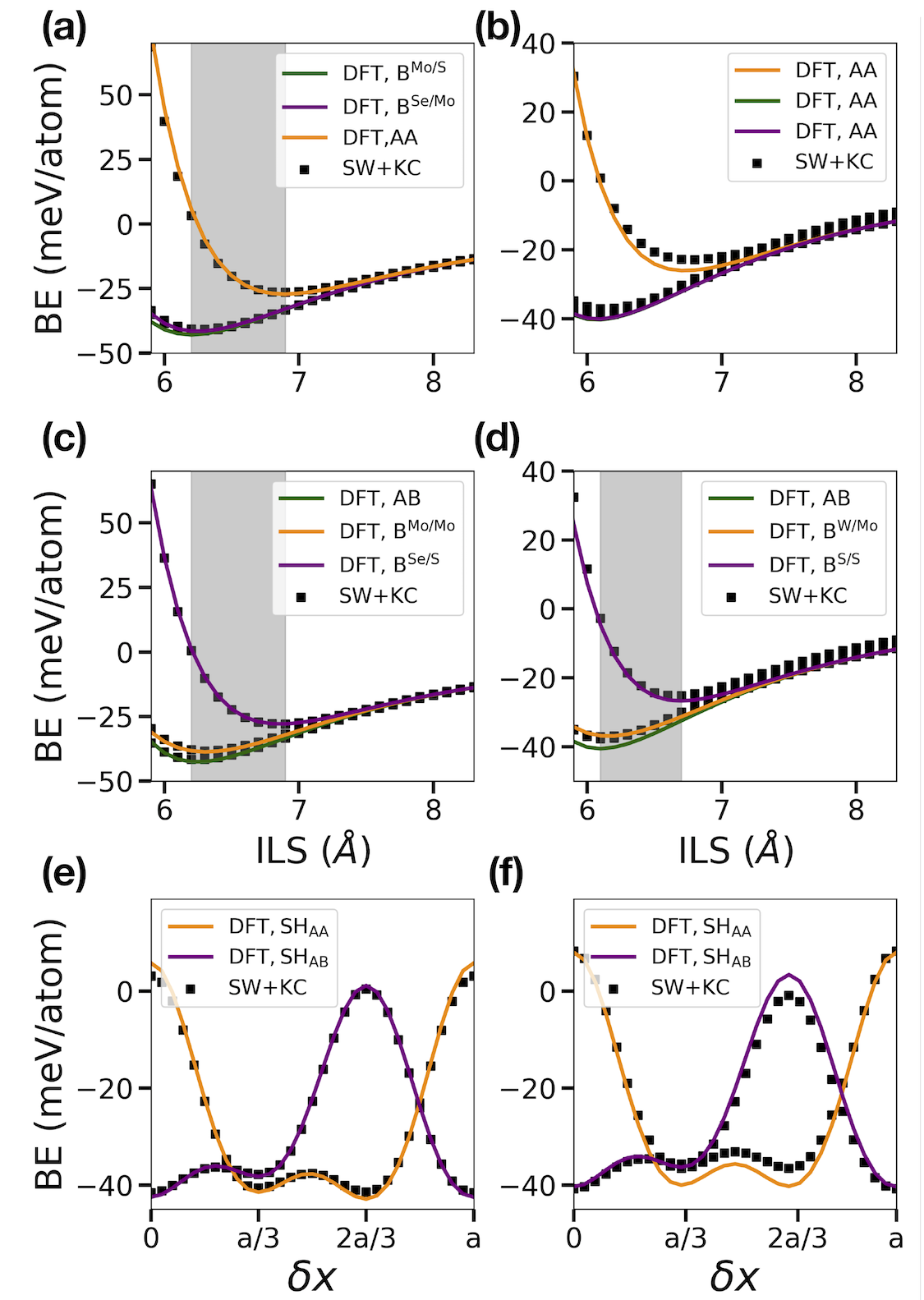}

 \caption{\label{figH}
 (a) and (c) ((b) and (d)) BE computed within DFT and using 
  SW+KC as a function of
  interlayer spacing (ILS) for different stackings in the heterostructure 
  MoS$_2$/MoSe$_2$ (MoS$_2$/WS$_2$). 
  The shaded region marks the range of possible interlayer spacings in the heterostructure.
 (e) ((f)) BE computed within DFT and using SW+KC as a function of 
 shear, $\mathrm{SH}_\mathrm{AA}$ and $\mathrm{SH}_\mathrm{AB}$, for MoS$_2$/MoSe$_2$ 
 (MoS$_2$/WS$_2$)
 The ILS here is fixed to that of the minimum ILS among the high-symmetry stackings.
    }
\end{figure}

We apply the mixing rule to simulate the MoS$_2$/MoSe$_2$ 
and MoS$_2$/WS$_2$ heterostructure.
MoS$_2$ and MoSe$_2$ are not lattice matched. For the purpose of 
comparison, we strain the layers to their average lattice constants in the DFT and forcefield
calculations. 
We compute the BE as a function of interlayer spacing and shear as shown in Fig. \ref{figH} 
for MoS$_2$/MoSe$_2$ and MoS$_2$/WS$_2$ heterostructure. 
AA stacking in the context of heterostructures implies transition metal 
(chalcogen) atom in the top layer is above
a transition metal (chalcogen) atom in the bottom layer. AB stacking implies 
transition metal (chalcogen) atom in the top layer is above
a chalcogen (transition metal) atom in the bottom layer.
The BE computed using SW+KC is in good agreement with DFT (Fig. \ref{figH}).   


\section{Moir\'e superlattices}

\begin{figure}
  \includegraphics[scale=0.67]{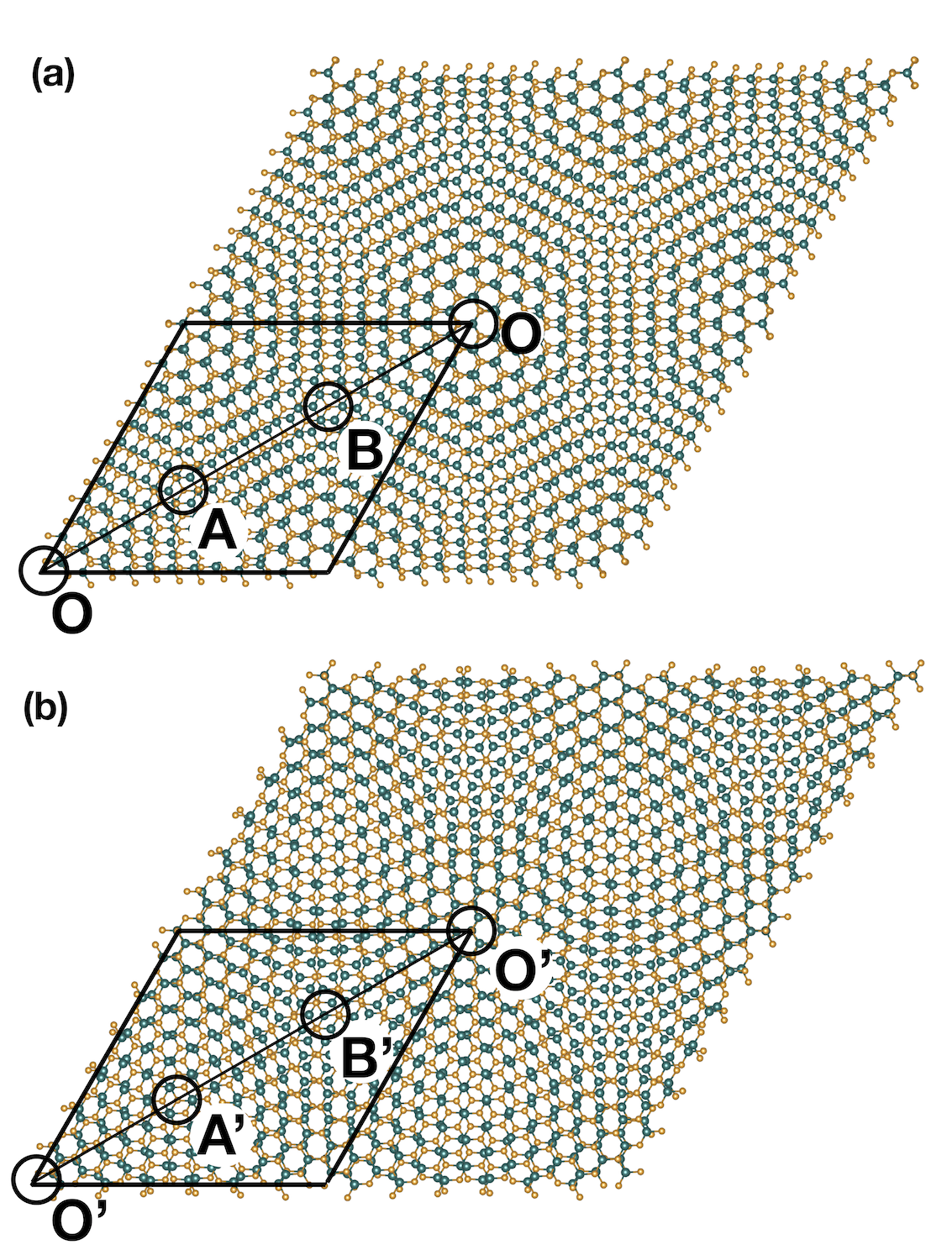}

\caption{\label{fig7}
  (a) and (b) Moir\'e superlattice (MSL) obtained on twisting bilayer MoS$_2$ by 5.1$^\circ$ and 
  54.9$^\circ$, respectively. Marked with circles are the high-symmetry stackings in the MSL. 
}

\end{figure}

\begin{figure*}
  \includegraphics[scale=1.8]{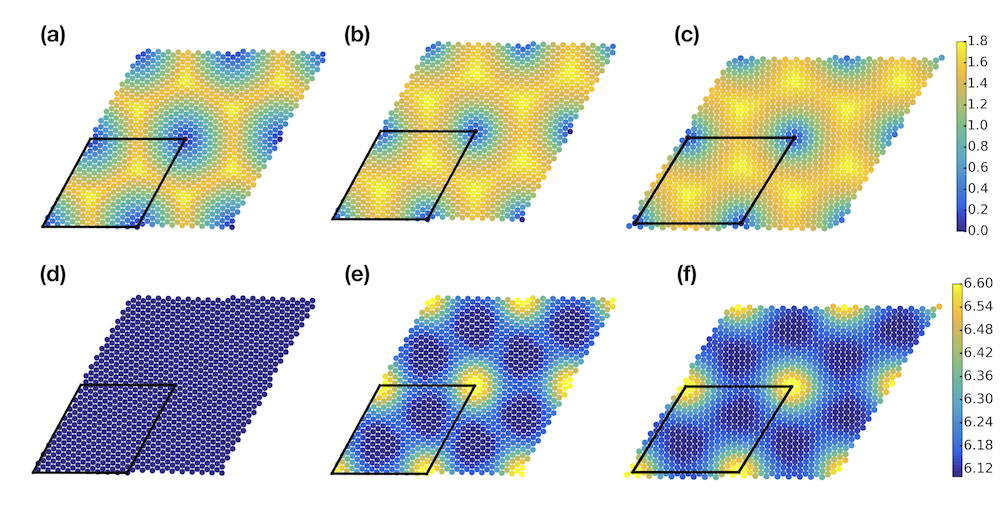}
  \caption{\label{fig8}
  Distribution of interlayer spacings (ILS) and order-parameter (OP) in the MSL. 
  (a) and (d) OP and ILS distribution in the rigidly-twisted bilayer MoS$_2$. The twist angle 
  here is 3.5$^\circ$. Black lines mark the MSL.
  (b) and (e) ((c) and (f)) Distribution of OP and ILS in the MSL relaxed using 
   DFT (SW+KC), respectively. The normals are taken into account 
   in the interlayer KC interaction.
}
\end{figure*}

Twisted bilayer TMDs form distinct moir\'e patterns for twist angles 
close to 0$^\circ$ and 60$^\circ$ \cite{twist.Naik, NL.Huang, AN.Puretzky}. 
Fig. \ref{fig7} shows the MSL for 
5.1$^\circ$ (M$^{5.1}$) and 54.9$^\circ$ (M$^{54.9}$) twists. 
The MSL is composed of the 
various high-symmetry stackings (Fig. \ref{fig1})\cite{twist.Naik}. 
The regions O, A and B are AA, 
B$^\mathrm{X/M}$ and B$^\mathrm{M/X}$ respectively. The regions O', A' and B' are AB,
B$^\mathrm{M/M}$ and B$^\mathrm{X/X}$ respectively.
Relaxing these moir\'e patterns in DFT starting 
from the rigidly twisted structure leads to significant in-plane and out-of-plane 
displacements of the atoms \cite{twist.Naik}. 
Because of large bending rigidity \cite{Maity_PRB_2018}, the out-of-plane displacements 
smoothly vary across the MSL surface. 
The ILS is largest for the AA and B$^\mathrm{S/S}$ stacking regions as shown in 
Fig. \ref{fig8} (e) for the case of 3.5$^\circ$ twisted bilayer MoS$_2$.
We also use the SW+KC to perform the relaxations of the same MSL, 
starting with the rigidly twisted structure. The ILS distribution obtained using 
the forcefield approach (Fig. \ref{fig8} (f)) is in good agreement with the DFT results.  
We have performed these relaxation using KC-n, ie. including
normals in the interlayer interaction. 

The undulations in each of the layers is smooth.
The out-of-plane displacements of each layer, while significant, vary over a large area.
We thus find that the normals do not 
deviate significantly from $\hat{z}$ for small or large twist-angles. The distribution 
of $\phi$ in KC-n relaxed structures is shown in Fig. \ref{figPhi} for a relatively 
large (7.3$^\circ$) and small (1$^\circ$) twist angle MSL. 
For angles greater than 7.3$^\circ$
and smaller than 52.7$^\circ$, the out-of-plane displacements are small. 
$\phi$ takes a maximum value of 
1.2$^\circ$ in Fig. \ref{figPhi}, and cos(1.2$^\circ$) $\approx$ 1. 
Furthermore, we find the energy computed using 
KC-n and KC-z for these angles differ by less than 0.01 meV/atom. 
KC-z is thus sufficient to capture relaxations in moir\'e superlattices.
We also compare the forces in the MSL computed using DFT, KC-n and KC-z for unrelaxed 
7.3$^\circ$ twisted bilayer MoS$_2$ in Fig. \ref{figForces}. 
The out-of-plane forces are in 
excellent agreement between DFT and the proposed forcefields. 

\begin{figure*}
  \includegraphics[scale=1.8]{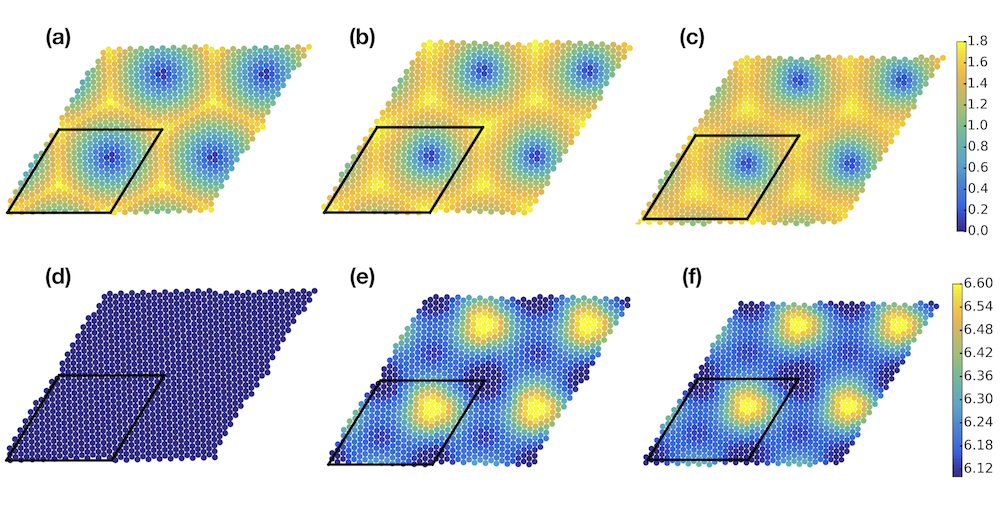}
  \caption{\label{fig9}
  Distribution of interlayer spacings (ILS) and order-parameter (OP) in the MSL.
  (a) and (d) OP and ILS distribution in the rigidly-twisted bilayer MoS$_2$. The twist angle
  here is 56.5$^\circ$.
  (b) and (e) ((c) and (f)) Distribution of OP and ILS in the MSL relaxed using
   DFT (SW+KC), respectively. The normals are taken into account
   in the interlayer interaction.
}
\end{figure*}

\begin{figure}
  \includegraphics[scale=1.8]{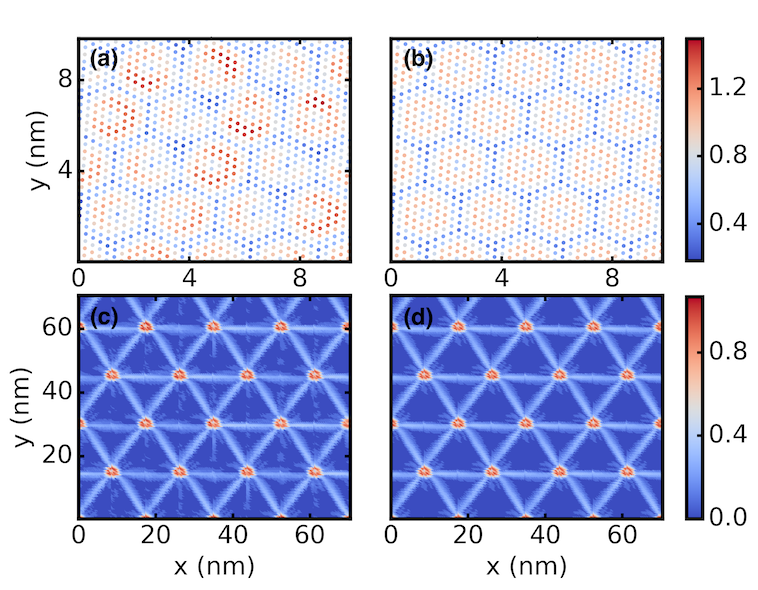}
  \caption{\label{figPhi}
  Distribution of the angular deviation, $\phi$ of the local unit-normal from the global $\hat{z}$ 
  direction (i.e. $n\cdot\hat{z}=cos(\phi)$) for $\mathrm{M}_1$ atoms in the MSL.   
  (a) and (b) $\phi$ computed for 7.3$^\circ$ twisted bilayer MoS$_2$, relaxed 
   using KC-n and KC-z, respectively.
  (c) and (d) $\phi$ computed for 1.0$^\circ$ twisted bilayer MoS$_2$, relaxed
   using KC-n and KC-z, respectively. The relaxations are performed using SW+KC. The colorbar shows the limits of the angular deviation distribution. 
}
\end{figure}

\begin{figure*}
  \includegraphics[scale=1.6]{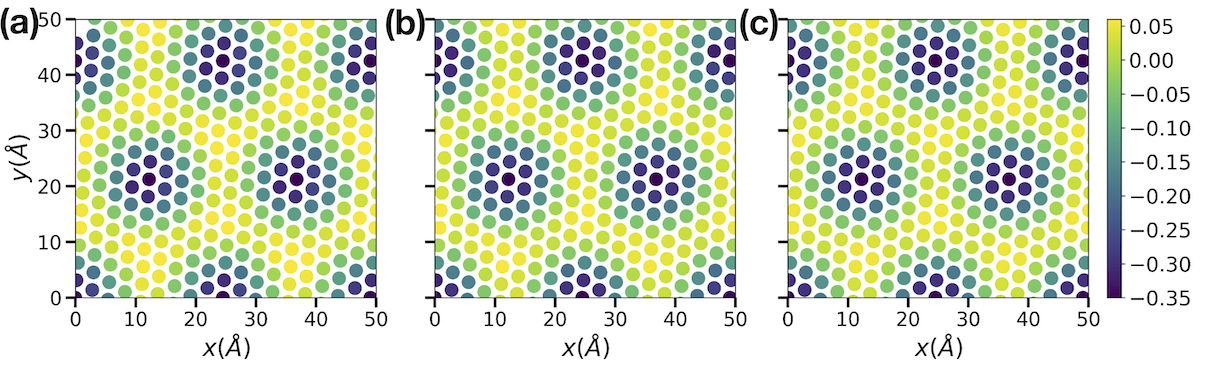}
  \caption{\label{figForces}
  Distribution of the out-of-plane component of forces, $f_z$ (in eV/\AA), 
  in 7.3$^\circ$ rigidly twisted bilayer MoS$_2$ (unrelaxed). $f_z$ is shown for the S$_2$ atom type in the bottom layer.
  (a) Computed using vdW corrected DFT.
  (b) and (c) Computed using
   SW+KC-n and 
   SW+KC-z, respectively. 
}
\end{figure*}


The in-plane displacements of the MoS$_2$ units in the moir\'e pattern 
lead to an increase in the area of 
low-energy stacking with respect to the rigidly twisted MSL \cite{twist.Naik}. 
The stackings in the 
 relaxed structure can be identified by means of an order-parameter (OP). The OP 
is defined \cite{twist.Naik} as the shortest displacement vector that 
takes any given stacking in the moir\'e pattern 
to the highest energy stacking. 
We define $\vec{u}$ for the moir\'e pattern formed by twist angles close to 
0$^\circ$ and $\vec{v}$ for those formed by twist angles close to
60$^\circ$. Hence $\vec{u}$ is the shortest displacement vector that takes any given stacking  
in M$^{3.5}$ to AA stacking and 
$\vec{v}$ is the shortest displacement vector that takes any given stacking  
in M$^{56.5}$ to B$^\mathrm{S/S}$ \cite{twist.Naik}. 
Fig. \ref{fig8} (a) shows the OP distribution for the 
rigidly twisted MSL, M$^{3.5}$. 
The distribution of the OP in the DFT relaxed MSL (Fig. \ref{fig8} (b)) is 
in good agreement with that in the SW+KC$^\mathrm{S-S}_\mathrm{Mo-S}$ relaxed MSL (Fig. \ref{fig8} (c)). 
This indicates
that the registry of atoms are similar in the MSL relaxed within DFT and the forcefield.
OP and ILS using the other set of parameters, KC$^\mathrm{S-S}_\mathrm{Mo-Mo}$ and 
KC$^\mathrm{S-S}_\mathrm{Mo-S}$,
are also in good agreement with DFT (see supporting information).

\begin{figure}
  \includegraphics[scale=0.52]{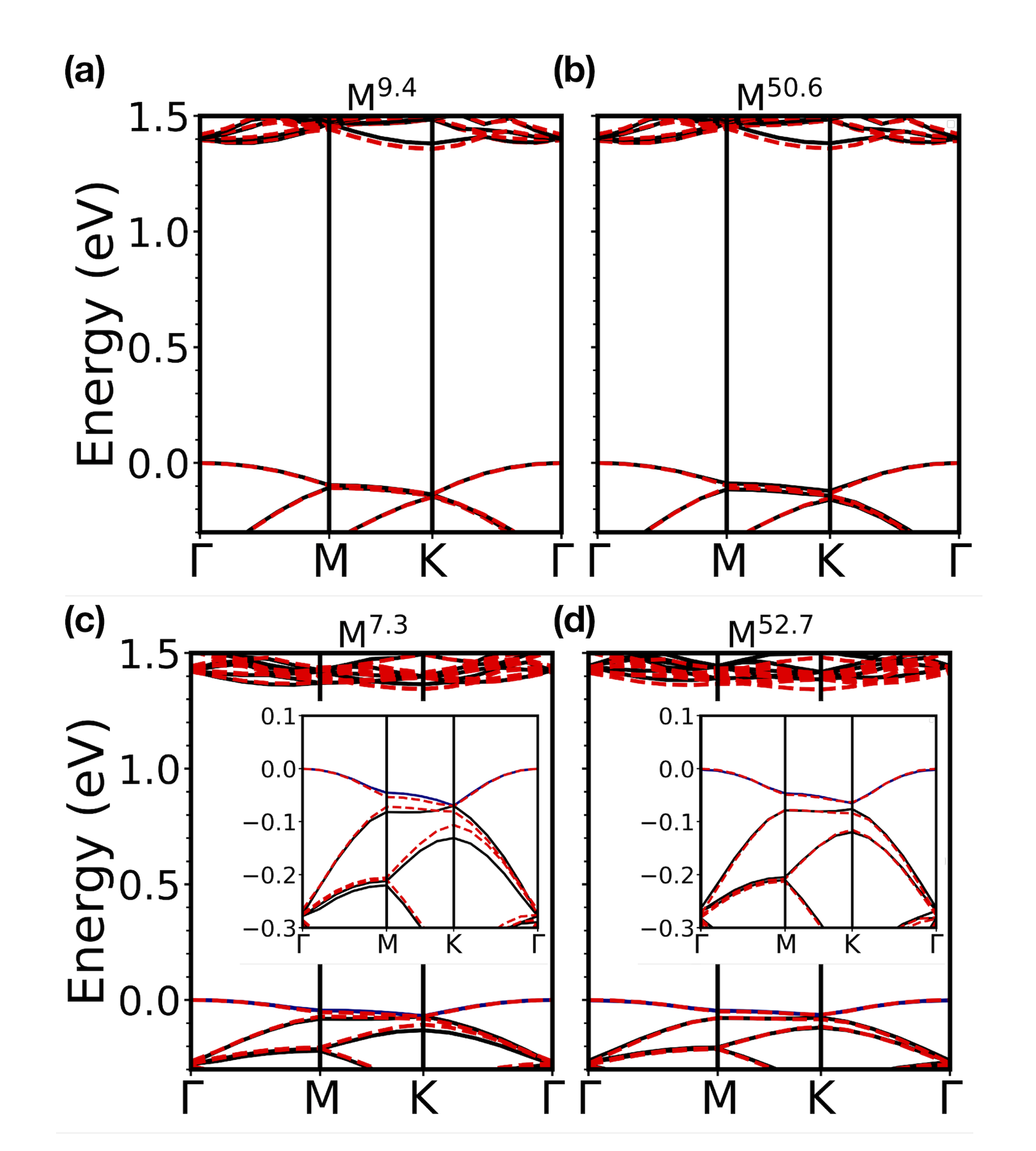}

 \caption{\label{fig10}
 (a) and (b) Electronic band structure of 9.4$^\circ$ and 
 50.6$^\circ$ twisted bilayer MoS$_2$, respectively. 
 (c) and (d) Band structure of 7.3$^\circ$ and 52.7$^\circ$ twisted bilayer MoS$_2$, respectively.
 Solid lines represent band structure computed using the DFT relaxed structure. 
 Dashed lines represent the 
 band structure computed using the forcefield (SW+KC) relaxed structure.
 The inset shows an enlarged plot of the valence bands in M$^{7.3}$ and M$^{52.7}$. 
 The flat band is marked in blue.
}
\end{figure}

The electronic structure of MSLs in bilayers \cite{twist.Naik} 
as well as heterostructures \cite{arxiv.Mac}
has been shown to host flat bands close to the valence band edge. We demonstrate that 
the electronic structure calculations can replace the time-consuming relaxation steps in DFT 
with the forcefield relaxations. 
To this effect, we compute the band structure of 7.3$^\circ$, 9.4$^\circ$, 
50.6$^\circ$, 52.7$^\circ$ twisted MSL using the DFT relaxed structure and the structure from 
SW+KC relaxations. We find that these band structures are in 
good agreement as shown in Fig. \ref{fig10}. 

In conclusion, we parametrize the KC potential for multilayer TMDs and 
propose a mixing rule to extend the parameter to TMD heterostructures. 
The parametrization is performed to fit the BE landscape computed using 
van der Waals corrected DFT. We show that the forcefield can be used 
to simulate structural transformations in MSLs of twisted bilayer MoS$_2$. These structural
reconstructions are in good agreement with those computed using DFT. Furthermore, 
the electronic band structure computed using the forcefield relaxed structure is in good 
agreement with band structure computed using the DFT relaxed structure. This indicates that 
the computationally expensive DFT relaxation steps can be replaced by the proposed forcefield 
method. Apart from predicting the structure of MSLs, several tribological properties can be 
studied using these KC potentials. The lubricity of the various bilayers and bilayer heterostructures 
can be computed. Moreover, the dependence of lubricity on the twist angle between the bilayers 
can be explored. The effect of finite temperature on the tribological 
properties and on solitons in the MSLs 
can also be studied with molecular dynamics simulations using these potentials.



\begin{acknowledgement}
We thank the Supercomputer Education and Research Centre (SERC) at IISc 
for providing the computational facilities.
\end{acknowledgement}

\begin{suppinfo}
Supporting Information. Verification
of computation of normals with peridically modulated MoS2 sheet; alternate parameter sets
KC$^\mathrm{X-X}_\mathrm{M-M}$ and KC$^\mathrm{X-X}_\mathrm{M-X}$; 
performance of the alternate parameter sets; table containing number of 
atoms in the MSL and MSL dimension;
ILS, OP and electronic band structure
of the MSL using SW+KC$^\mathrm{S-S}_\mathrm{Mo-Mo}$ for twisted bilayer MoS$_2$; 
refitted parameters with the 
taper function for KC$^\mathrm{X-X}_\mathrm{M-M}$.
\end{suppinfo}

\providecommand{\latin}[1]{#1}
\makeatletter
\providecommand{\doi}
  {\begingroup\let\do\@makeother\dospecials
  \catcode`\{=1 \catcode`\}=2 \doi@aux}
\providecommand{\doi@aux}[1]{\endgroup\texttt{#1}}
\makeatother
\providecommand*\mcitethebibliography{\thebibliography}
\csname @ifundefined\endcsname{endmcitethebibliography}
  {\let\endmcitethebibliography\endthebibliography}{}

\end{document}